\documentclass[12pt]{iopart}

\begin{document}

\title[On the measurement of $B(E2)$ using intermediate-energy Coulomb excitation]{On the measurement of $B(E2, 0^+_1 \rightarrow 2^+_1)$ 
using intermediate-energy Coulomb excitation}

\author{F.~Delaunay$^{1,2}$ and F.~M.~Nunes$^{2,3}$}

\address{$^1$ Laboratoire de Physique Corpusculaire, ENSICAEN, Universit\'e de Caen, CNRS/IN2P3, 14050 Caen, France}

\address{$^2$ National Superconducting Cyclotron Laboratory, 
Michigan State University, East Lansing, Michigan 48824, USA}

\address{$^3$ Department of Physics and Astronomy, 
Michigan State University, East Lansing, Michigan 48824, USA}

\ead{delaunay@lpccaen.in2p3.fr}

\date{\today}

\begin{abstract}
Coulomb excitation is a standard method used to extract quadrupole excitation
strengths of even-even nuclei. In typical analyses the reaction is assumed to be one-step,
Coulomb only, and is treated within a semi-classical model. In this work,
fully-quantal coupled-channel calculations are performed for three test cases
in order to determine the importance of multi-step effects, nuclear contributions,
feeding from other states and corrections to the semi-classical approximation.
We study the excitation of $^{30}$S, $^{58}$Ni and
$^{78}$Kr on $^{197}$Au at $\approx$ 50 AMeV. We find that nuclear
effects may contribute more than 10\% and that feeding contributions can be larger
than 15\%.
These corrections do not alter significantly the published $B(E2)$ values, however
an additional theoretical error of up to 13\% should be added to the experimental
uncertainty if the semi-classical model is used. This theoretical error is reduced 
to less than 7\% when performing a quantal coupled-channel analysis.
\end{abstract}

\submitto{\JPG}

\maketitle

Nuclear collectivity of an even-even nucleus is closely related to its 
quadrupole electric reduced transition probability $B(E2, 0^+_1 \rightarrow 2^+_1)$. This strength
can be determined
experimentally by measuring either the lifetime or the Coulomb excitation
cross section of the $2^+_1$ excited state \cite{cook,coulex-rev}.
Originally, the Coulomb excitation technique (referred to as ``Coulex'' in the following)
was used to measure properties of the target (e.g. \cite{shaw,ahmad}). The sub-Coulomb energies
at which the reaction took place ensured a nuclear-free measurement.
In  the last decade, Coulex has been expanded to intermediate energies with
the aim of studying unstable nuclei \cite{coulex-rev}. In this case the nucleus of interest is 
the beam particle and a heavy target is used to produce the virtual photons. 
The reaction takes place at high enough energy to inhibit multi-step effects 
and data  is taken only at very forward angles, where one expects to be free 
from nuclear interference. This method has enabled accurate measurement of the $B(E2)$
of a large variety of systems
\cite{motobayashi,scheit,ibbotson,cottle1,cottle2,gade1,yurkewicz1,yurkewicz2,dinca,gade2,church,banu,perru}.

A systematic comparison between the intermediate-energy Coulex method and
the lifetime method showed that there is consistency between the two techniques \cite{cook}.
Moreover, the accuracy of the $B(E2)$ strengths extracted through Coulex
is comparable to that from the lifetime measurements \cite{cook}. The work of Cook 
{\it et al.} \cite{cook} focused on the experimental accuracy but did not consider
the uncertainties due to the approximations in the Coulex theory used 
to connect cross sections and electric strengths. This is exactly the focus of 
the present study.

One way to analyse these unstable beam Coulex experiments is 
using the semi-classical model of Alder and Winther \cite{aw2}. 
It conveniently provides a linear relation between the 
Coulex cross section and the reduced transition probability.
The approximations in the Alder and Winther theory \cite{aw2} are three fold:
1) the straight-line semi-classical approximation; 2) the excitation is a one-step
process; 3) it is purely Coulomb. The straight-line approximation is partially corrected
within Alder and Winther \cite{aw2}. The second point is not so straight-forward.
Most of the nuclei studied through this technique exhibit
large collectivity and thus have other excited states that 
are strongly coupled to either the ground state or the first excited $2^+$ state. 
Even if it is generally assumed that the cross sections to these other excited
states are small at intermediate beam energies, multi-step mechanisms and interferences
can distort the desired result. In order to solidify the reliability of the intermediate-energy
Coulex method, it is important to evaluate the uncertainties 
coming from the one-step approximation. Finally, nuclear contributions need to be 
consistently included in the calculations so that Coulomb-nuclear interference is
correctly accounted for. The inclusion of nuclear effects in realistic quantum
coupled-channel calculations may enhance multi-step effects.

As mentioned above, intermediate-energy Coulex relies on 
restricting the scattering angles taken into account for integrating the cross sections
to a range corresponding to impact parameters larger than the sum of the target and projectile 
radii.
A detailed study of the sensitivity to the impact parameter cut was performed on the 
$^{46}$Ar data \cite{gade1} and results validate the procedure. In cases where low statistics
forces the inclusion of a wider angular range, nuclear effects have been estimated
with quantum distorted wave calculations to be of the order of 6\% \cite{yurkewicz1,church}.
However, this value should not be taken as definitive, since, as we shall see,
the nuclear contribution depends strongly on the particular analysis considered.

Another problem that is considered when analyzing Coulex data
is the possibility of feeding: the reaction process excites high lying states 
that could then decay to the $2^+_1$ state, producing an enhanced 
$2^+_1 \rightarrow 0^+_1$ signal. In most of the studies, estimates of feeding predict
it to be unimportant (e.g. \cite{dinca,church}) mostly because at intermediate energies
the relative cross sections to higher spin states are small and
larger excited states are hindered compared to the lower transitions.
Nevertheless there have been cases where feeding needs to be carefully considered 
before a reliable strength is extracted \cite{yurkewicz2,cottle1}. 
In intermediate-energy Coulex experiments, the statistics is often low 
and the efficiency of the $\gamma$-ray detectors is limited such that a $\gamma$ peak
for a feeding transition is rarely seen \cite{cottle1}. Feeding corrections are based
on theoretical estimates \cite{aw2} and subtracted from the $2_1^+$ cross section, before extracting
the $B(E2)$ strength.

Intermediate-energy Coulex has been applied mostly to intermediate
mass nuclei bound by a few MeV but as beam intensities improve, it will be applied
to more exotic systems.
The loosely bound nature of unstable nuclei has modified many of the traditional
views of nuclear reactions. For example, when the exotic nucleus has an extended
tail in its wave-functions, one finds nuclear contributions at impact parameters much
larger than the sum of the target and projectile radii \cite{nunes98}. In addition,
due to the proximity to the continuum, multi-step breakup effects need to be considered \cite{nunes99}.
A systematic study of nuclear interference in the Coulomb dissociation 
of halo nuclei has shown large nuclear effects, even in the forward
angular regions considered safe for Coulomb experiments \cite{hussein06}. 
A recent comprehensive study of the Coulex of $^{11}$Be for extraction of the $B(E1)$
between the two bound states validates the Coulex method across a wide range of beam energies,
provided all these effects are taken into account in the theoretical model \cite{summers}.
For the $B(E2)$ of intermediate mass nuclei, it is important to solidify 
the theoretical methods used at present before these new dripline challenges can be faced.

\begin{table}[b!]
\caption{Information on the intermediate-energy Coulex 
experiments considered here. For each case we give the beam laboratory energy, 
the maximum centre-of-mass angle for cross-section integration, the corresponding cross section 
for the $2_1^+$ state and the $B(E2, 0^+_1 \rightarrow 2^+_1)$ value extracted through 
Winther and Alder's theory \cite{aw2}.}
\begin{center}
\begin{tabular}{ccccc}
\hline
\hline
Nucleus & Energy & $\theta_{CM}^{max}$ & $\sigma_{2^+_1}$ & $B(E2,0^+_1 \rightarrow 2^+_1)$ \\
& (AMeV) & (deg.) & (mb) & ($e^2$fm$^4$) \\
\hline
$^{30}$S & 35.7 & 4.56 & 39.6(3.8) & 350(33) \\
$^{58}$Ni & 72.4 & 4.26 & 175(36) & 707(145) \\
$^{78}$Kr & 57.4 & 4.24 & 1124(133) & 6244(738) \\
\hline
\hline
\end{tabular}
\end{center}
\label{TableExp}
\end{table}

In this work we perform fully-quantum coupled-channel calculations for three
test cases that have been measured by intermediate-energy Coulex:
$^{30}$S, $^{58}$Ni and $^{78}$Kr. These three test cases span a variety of
physical situations. The first, $^{30}$S, corresponds to a very short lived isotope, 
two nucleons away from the proton dripline, with only
a few excited states. The Coulex of $^{30}$S was measured at 35.7 AMeV on $^{197}$Au
\cite{cottle2}. The second, $^{58}$Ni is less exotic, contains very strong transitions
to higher energy states and therefore has an important feeding correction. It has been
measured several times before and we consider here the experiment 
at 72.4 AMeV on $^{197}$Au \cite{yurkewicz2}. The third, $^{78}$Kr,
has a very small $2^+_1$ excitation energy, and consequently a very large $B(E2)$.
Here we consider a recent measurement  at $57.4$ AMeV on a  $^{197}$Au target 
\cite{gade2}. Experimental details for these experiments are summarized in 
Table \ref{TableExp}, where we include the $B(E2)$ extracted in the corresponding
studies using the first order semi-classical theory \cite{aw2}.

\begin{table}[t!]
\caption{Spin, parity and excitation energy for all the states included in the 
coupled-channel calculations. The $0^+$ assignment for the 3.666 MeV state in 
$^{30}$S is based on a comparison of the experimental spectrum with the spectrum 
of the mirror nucleus and a shell-model calculation for the $A=30$ isobars.}
\begin{center}
\begin{tabular}{cc|cc|cc}
\hline
\hline
\multicolumn{2}{c|}{$^{30}$S} & \multicolumn{2}{c|}{$^{58}$Ni} & \multicolumn{2}{c}{$^{78}$Kr}\\
\hline
$E$ (MeV) & $J^\pi_n$ & $E$ (MeV) & $J^\pi_n$ & $E$ (MeV) & $J^\pi_n$ \\
\hline
0         & 0$^+_1$ & 0         & 0$^+_1$ & 0         & 0$^+_1$ \\
2.211 & 2$^+_1$ & 1.454 & 2$^+_1$ & 0.455 & 2$^+_1$ \\
3.403 & 2$^+_2$ & 2.459 & 4$^+_1$ & 1.017 & 0$^+_2$ \\
3.666 & (0$^+_2$) & 2.775 & 2$^+_2$ & 1.119 & 4$^+_1$ \\
           &                  & 3.038 & 2$^+_3$ & 1.148 & 2$^+_2$ \\
           &                  & 3.263 & 2$^+_4$ &            &                  \\
\hline
\hline
\end{tabular}
\end{center}
\label{TableStates}
\end{table}


We have investigated the spectra of these nuclei in detail and isolated
the states that can affect the reaction mechanism. These are summarized
in Table \ref{TableStates}. For the two heavier cases, the spectra are well known.
However, for $^{30}$S, the spin and parity of the 3.666 MeV state are undetermined. 
We have assumed they are $0^+$ by comparison with the level scheme of the $^{30}$Si 
mirror nucleus \cite{NNDC} and a shell-model calculation for the $T=1$ states of the $A=30$ isobars \cite{BrownWeb}. 
As it can feed into the $2^+_1$ state, it needs to be included in the calculations.

States with unnatural parity (e.g. $1^+$ and $3^+$ states) were not included since they would 
decay to $2^+$ and $0^+$ states by magnetic transitions which are not implemented in our
coupled-channel calculations. In the energy range of interest, there is one $1^+$ state
in $^{30}$S, one $1^+$ state in $^{58}$Ni and none in $^{78}$Kr.
Generally, magnetic transitions in Coulex are much weaker than the electric ones \cite{rmp}.

\begin{table}[t!]
\caption{Transitions included in the calculations: spins and parities of the states involved in 
each transition, the corresponding halflives $T_{1/2}$, branching ratios $I_{\gamma}$, 
the reduced transition matrix elements $M(E\lambda)$ and deformation lengths $\delta=\beta R$. 
Unless otherwise noted, the spins, parities, halflives and branching ratios were taken 
from the NNDC database \cite{NNDC}.}
\begin{center}
\begin{tabular}{ccccccc}
\hline
\hline
& $J^\pi_{n,1}$ & $J^\pi_{n,2}$ & $T_{1/2}$ & $I_\gamma$ & $M(E\lambda)$ & $\beta R$ \\
& & & & & ($e$fm$^\lambda$) & (fm) \\
\hline
$^{30}$S & $2^+_1$ & $0^+_1$ &             & 1.00 & 18.71 $^a$ & 1.314 $^a$\\
                  & $2^+_2$ & $0^+_1$ & 115 fs & 0.20 &  3.29          & 0.231 \\
                  & $2^+_2$ & $2^+_1$ & 115 fs & 0.80 & 90.51         & 2.842 \\
                  & $0^+_2$ & $2^+_1$ & 1 ps $^b$ & 1.00 &  9.31          & 0.292 \\
\hline
$^{58}$Ni & $2^+_1$ & $0^+_1$ &           & 1.00 & 26.59 $^a$ & 0.856 $^a$ \\
                   & $4^+_1$ & $0^+_1$ & 970 fs $^b$ & 2$\times$10$^{-8}$ $^c$ & 481.08 & 0.718 \\
                   & $4^+_1$ & $2^+_1$ & 970 fs $^b$ & 1.00 & 71.56 & 1.031 \\
                   & $2^+_2$ & $0^+_1$ & 0.38 ps & 4.3$\times$10$^{-2}$ & 1.40 & 0.045 \\
                   & $2^+_2$ & $2^+_1$ & 0.38 ps & 0.96 & 42.13 & 0.607 \\
                   & $2^+_2$ & $4^+_1$ & 0.38 ps & 5.7$\times$10$^{-4}$ & 36.75 & 0.395 \\
                   & $2^+_3$ & $0^+_1$ & 52 fs & 0.40 & 9.18 & 0.296 \\
                   & $2^+_3$ & $2^+_1$ & 52 fs & 0.58 & 56.58 & 0.815 \\
                   & $2^+_3$ & $4^+_1$ & 52 fs & 2.9$\times$10$^{-3}$ & 49.62 & 0.533 \\
                   & $2^+_3$ & $2^+_2$ & 52 fs & 9.9$\times$10$^{-3}$ & 72.18 $^d$ & 0.775 $^d$ \\
                   & $2^+_4$ & $0^+_1$ & 35 fs & 0.59 & 11.41 & 0.367 \\
                   & $2^+_4$ & $2^+_1$ & 35 fs & 0.39 & 40.45 & 0.583 \\
                   & $2^+_4$ & $4^+_1$ & 35 fs & 1.0$\times$10$^{-2}$ & 49.35 & 0.530 \\
                   & $2^+_4$ & $2^+_2$ & 35 fs & 1.8$\times$10$^{-3}$ & 72.18 & 1.040 \\
\hline
$^{78}$Kr & $2^+_1$ & $0^+_1$ &           & 1.00 & 79.02 $^a$ & 1.793 $^a$ \\
                   & $0^+_2$ & $2^+_1$ & 5.8 ps $^e$ & 1.00 & 41.89 $^e$ & 0.425 $^e$ \\
                   & $4^+_1$ & $2^+_1$ & 2.5 ps & 1.00 & 125.68 & 1.275 \\
                   & $2^+_2$ & $0^+_1$ & 3.7 ps & 0.39 & 12.21 & 0.277 \\
                   & $2^+_2$ & $2^+_1$ & 3.7 ps & 0.61 & 54.29 & 0.551 \\
\hline
\hline
\multicolumn{7}{l}{$^a$ using $B(E2,0^+_1 \rightarrow 2^+_1)$ measured by Coulex
\cite{cottle2,yurkewicz2,gade2}}\\
\multicolumn{7}{l}{$^b$ experimental lower value} \\
\multicolumn{7}{l}{$^c$ $I_\gamma$ of the $4^+_1 \rightarrow 0^+_1$ transition in $^{60}$Ni}\\
\multicolumn{7}{l}{$^d$ from the $2^+_4 \rightarrow 2^+_2$ transition in $^{58}$Ni}\\
\multicolumn{7}{l}{$^e$ assuming $B(E2, 0^+_2 \rightarrow 2^+_1) = B(E2,4^+_1 \rightarrow 2^+_1)$}\\
\end{tabular}
\end{center}
\label{TableTransitions}
\end{table}

A very large amount of transitions is possible between the excited states
listed in Table \ref{TableStates}. 
In Table \ref{TableTransitions} we list all the transitions taken
into account in our coupled-channel calculations. We also provide halflives and branching
ratios \cite{NNDC} from which we determined the $B(E\lambda)$. 
Reduced matrix elements $M(E\lambda)$  
were evaluated directly from $B(E\lambda)$:
$$
M(E\lambda,I_1 \rightarrow I_2) = \sqrt{(2I_1+1)B(E\lambda,I_1 \rightarrow I_2)}.
$$
For the $0_1^+ \rightarrow 2_1^+$ transitions,  we used $B(E2)$ directly from
the Coulex experiments data under scrutiny \cite{cottle2,yurkewicz2,gade2}.
We have not included transitions with reduced matrix elements $M(E\lambda)$ 
two orders of magnitude lower than the main $0^+_1 \rightarrow 2^+_1$ transition. 
Since our coupled-channel calculations are restricted to electric couplings, transitions involving a
mixture of $M1$ and $E2$ transitions (such as $2^+ \rightarrow 2^+$ transitions) were assumed to be
100\% $E2$.
This is not always a good approximation. However, none of these transitions is involved in the
direct excitation of the $2^+_1$ states, but in feeding couplings.
Our aim was to study the effects of couplings, therefore we chose to keep the full strength of these
transitions. In our calculations we assumed all coupling amplitudes have the same sign. We have
checked this assumption and verified that uncertainties due to sign changes are negligible.

A specific remark is needed for $^{78}$Kr, as
the lifetime of the $0^+_2$ state is unknown.
This state, as well as the $4_1^+$ and $2^+_2$ states, lies at about twice the excitation
energy of the $2_1^+$ state. 
This suggests that these $0^+_2$, $2^+_2$ and $4^+_1$ states are quadrupole 2-phonon states, 
with the $2_1^+$ being the {\mbox 1-phonon} state. In a pure vibrational model, 
$B(E2,0^+_2 \rightarrow 2_1^+)=B(E2,4_1^+ \rightarrow 2_1^+)$ \cite{BohrMottelson2}. 
This is what we assumed in order to obtain the transition matrix element 
for the $0^+_2 \rightarrow 2_1^+$ transition in $^{78}$Kr.

As part of our aim is to check the importance of nuclear effects, nuclear couplings 
were also included in the calculations. We have assumed the matter deformation to be
the same as the charge deformation for all three cases. As these systems are not halo-like, this 
approximation should be adequate. The corresponding deformation lengths 
are also included in Table \ref{TableTransitions}. For a given transition, 
the nuclear and Coulomb couplings were considered in both directions, 
but reorientation couplings were not included, as would be the case in a vibrational model.
Optical model parameters were taken from elastic scattering studies.
For the $^{30}$S case, we used the optical model parameters for $^{40}$Ar + $^{208}$Pb 
at 44 AMeV \cite{Alamanos84}. For $^{58}$Ni and $^{78}$Kr on $^{197}$Au, we used the parameters 
from an elastic scattering study of $^{86}$Kr + $^{208}$Pb at 43 AMeV \cite{RousselChomaz88}. 
A value of $r_C=1.2$ fm was used for the Coulomb radius parameter.
Coupled-channel calculations were performed using the code FRESCO \cite{FRESCO}. 

\begin{table}[t!]
\caption{One-step (DWBA) and coupled-channel (CC) calculations including only the ground state and the $2_1^+$ state: comparison between cross sections integrated over the experimental angular
range for the full process with those from Coulomb only. All cross sections are in mb.}
\begin{center}
\begin{tabular}{l|r|r}
\hline
\hline
     & DWBA($2^+_1$) & CC($2^+_1$) \\
\hline
{$^{30}$S} Coul+nucl & 44.7 & 43.6 \\
{$^{30}$S} Coulomb & 40.6 & 40.6 \\
\hline
{$^{58}$Ni} Coul+nucl & 161.5 & 161.1 \\
{$^{58}$Ni} Coulomb & 183.8 & 182.7 \\
\hline
{$^{78}$Kr} Coul+nucl & 1052.6 & 1030.7 \\
{$^{78}$Kr} Coulomb & 1056.3 & 1036.0 \\
\hline
\hline
\end{tabular}
\end{center}
\label{TableResults1}
\end{table}

For an accurate calculation of the Coulex cross section at forward angles 
one needs to be very careful with convergence.
Partial waves up to $L_{max}=3000-6000$ and a radial integration up to $R_{max}=300$ fm were needed
in order to get a converged integrated cross section within the experimental angular range.
Checks of the sensitivity of the $2_1^+$ excitation cross section to the optical model 
parameters were performed and we estimate an uncertainty smaller than 5\%.
Our calculations show negligible sensitivity to the Coulomb radius $r_C$.
It is important to note that the calculations we have performed are non-relativistic.
Relativistic kinematical effects have been studied within the context of the
first-order semi-classical theory \cite{bertulani}.
We have estimated the effect of relativistic kinematics by repeating the
calculations at a corrected beam energy
and obtained for all three cases modifications in the cross sections smaller than 5\%.
Therefore, if we neglect uncertainties in the transition strengths used,
the theoretical error on our cross sections should be smaller than 7\%.
This should be added to the experimental errors.

\begin{table}[t!]
\caption{Full coupled-channel calculations including all transitions specified 
in Table \ref{TableTransitions}. $I_{J_n^\pi \rightarrow 2^+_1}$ is the branching ratio for the 
$J_n^\pi \rightarrow 2_1^+$ transition \cite{NNDC}. $\sigma_{feed}$ are the cross sections for 
feeding into the $2_1^+$ state. The value for $\sigma_{feed}$ in bold is the sum of 
the cross sections contributing to $\sigma(2_1^+)$ for each case. }
\begin{center}
\begin{tabular}{c|c|ccc}
\hline
\hline
        &        & \multicolumn{3}{c}{Full CC} \\
Nucleus & State  &  $\sigma$ & $I_{J_n^\pi \rightarrow 2^+_1}$ &$\sigma_{feed}$ \\
        &        & (mb) & & (mb) \\
\hline
         & $2^+_1$ & 43.6 & & {\bf 45.5} \\
$^{30}$S & $2^+_2$ & 2.2 & 0.80  &1.8 \\
         & $0^+_2$ &0.1 & 1.00 & 0.1 \\
\hline
          & $2^+_1$ & 155.6 & & {\bf 188.3} \\
          & $4^+_1$ & 8.6 & 1.00  & 8.6 \\
$^{58}$Ni & $2^+_2$ & 1.9 & 0.96 & 1.8 \\
          & $2^+_3$ & 19.1 & 0.58 & 11.1 \\
          & $2^+_4$ & 29.0 & 0.39 & 11.3 \\
\hline
          & $2^+_1$  & 1013.9 & & {\bf 1041.9} \\
$^{78}$Kr & $0^+_2$  & 2.8 & 1.00 & 2.8 \\
          & $4^+_1$ & 8.0 & 1.00 & 8.0 \\
          & $2^+_2$ & 28.1 & 0.61 & 17.2 \\
\hline
\hline
\end{tabular}
\end{center}
\label{TableResults2}
\end{table}

Results including only the ground and $2^+_1$ states are presented 
in Table \ref{TableResults1}. We present 1-step (DWBA) and coupled-channel (CC)
calculations. Cross sections were integrated over the angular range of the particular
experiment. We performed calculations including both nuclear and Coulomb excitations
in the transition matrix elements and we compare them with the results obtained with only
Coulomb. In all three cases, $0_1^+ \leftrightarrow 2_1^+$ multi-step effects are small.
Nuclear effects are of the order of 10\%, 12\% and 0.4\% for
$^{30}$S, $^{58}$Ni and $^{78}$Kr, respectively. These are still within the experimental limits.

For $^{30}$S and $^{58}$Ni, nuclear effects are not negligible and indicate that the maximum angle
used in integrating the experimental cross section should be carefully chosen.
The contribution of the nuclear part of the interaction to the cross section is 
indeed very sensitive to this maximum angle. For $^{58}$Ni, our tests show that decreasing
the maximum centre-of-mass angle by only 0.5 degree cuts the relative nuclear contribution
by a factor 2.

One can also compare our Coulomb-only DWBA cross
sections to those predicted by the semi-classical model (see Table \ref{TableExp}).
From this comparison we find that the straight-line trajectory approximation alone
introduces an error in the cross section of 6\% at the most.

Full coupled-channel results including the states in Table \ref{TableStates} and transitions from
Table \ref{TableTransitions} are presented in Table \ref{TableResults2}. The cross sections
to individual states ($\sigma$) are multiplied by the branching ratios to the $2_1^+$ state
in order to get the feeding contributions. The sum of these and the Coulex $2^+_1$ cross section
gives the full cross section (in bold) to be compared to the experiment.
Feeding contributions are important in $^{58}$Ni ($\approx$ 17\%), as we already knew, but also in 
$^{30}$S ($\approx 5$\%).  For $^{58}$Ni, the feeding correction estimated in \cite{yurkewicz2} 
of $25 $ mb is significantly smaller than our prediction (33 mb), mainly because of the contribution
from the $4^+_1$ state which was omitted in the experimental estimation \cite{yurkewicz2}.
For $^{30}$S our theoretical cross section taking 
$B(E2, 0_1^+ \rightarrow 2_1^+)=350$ e$^2$fm$^2$ is above the upper limit of the experimental range.
A reduction of the $B(E2)$ value to 303 e$^2$fm$^4$ is necessary to get a theoretical 
cross section in agreement with the experimental result. Considering the $7$\% theoretical error,
this value is still compatible with the value extracted in \cite{cottle2}.
For $^{58}$Ni and $^{78}$Kr, the expected cross section is within 
$\approx 7$\% of the mean experimental value, i.e. within the experimental range.

In conclusion, full coupled-channel calculations for $^{30}$S, $^{58}$Ni and $^{78}$Kr Coulex
confirm the agreement between $B(E2)$ extracted through the lifetime and the
Coulex methods \cite{cook}. Our study shows that theoretical contributions to the
errors in the cross sections can be $\approx 13$\% if the first order semi-classical
theory of Alder and Winther is used \cite{aw2} but should be less than 7\% if
a full quantum coupled-channel calculation is performed and feeding is consistently
taken into account.

We thank Alexandra Gade for useful discussions and comments on earlier versions of the
manuscript. This work was supported by NSCL, Michigan State University,
and the National Science Foundation through grant PHY-0555893.

\end{document}